\documentclass[11pt]{article}
\usepackage{moriond,epsfig}

\def\lsim{\raise0.3ex\hbox{$<$\kern-0.75em\raise-1.1ex\hbox{$\sim$}}}
\def\gsim{\raise0.3ex\hbox{$>$\kern-0.75em\raise-1.1ex\hbox{$\sim$}}}

\def\be{\begin{equation}}
\def\ee{\end{equation}}
\def\bea{\begin{eqnarray}}
\def\eea{\end{eqnarray}}

\begin{document}
\vspace*{4cm}

\title{CONSTRAINTS FOR NUCLEAR GLUON DENSITIES FROM DIS DATA}

\author{
 K.J. Eskola$^{\rm a,b,}$,
 H. Honkanen$^{\rm a,b}$,
 V.J. Kolhinen$^{\rm a,b}$,
 \underline{C.A. Salgado}$^{\rm c}$.
}

\address{
{\em $^{\rm a}$ Department of Physics, University of Jyv\"askyl\"a,\\
P.O.Box 35, FIN-40351
Jyv\"askyl\"a, Finland\\}
{\em $^{\rm b}$ Helsinki Institute of Physics,\\
P.O.Box 64, FIN-00014 University of Helsinki, Finland\\}
{\em $^{\rm c}$ CERN, Theory Division, CH-1211 Geneva, Switzerland}
}

\maketitle\abstracts{
The $Q^2$ dependence of the ratios of nuclear structure functions  $F_2^A$
is studied
by performing QCD evolution of nuclear parton distribution functions.
The log $Q^2$ slope of these ratios is
very sensitive to the nuclear gluon distribution function.  Taking
different parametrizations, we show that the NMC data on the $Q^2$
dependence of $F_2^{\rm Sn}/F_2^{\rm C}$ rule out the case where
nuclear shadowing (suppression) of gluons at $x\sim 0.01$ is much
larger than the shadowing observed in the ratio $F_2^A/F_2^{\rm D}$.
We also take into account modifications to the DGLAP evolution by
including gluon fusion terms and see that the effect is small at
present energies, and, in any case, 
a strong gluon shadowing is not favored. The region studied ($x \sim
0.01$) is the most relevant for RHIC multiplicities.
}

The nuclear parton distribution functions (nPDFs) are an essential
ingredient in 
calculations of hard processes involving nuclei.  In
the framework of the QCD improved parton model it is possible to
extract the 
nPDFs from experiments of
DIS and to use them for other processes in $AB$ collisions. In this
framework, a set of nPDFs, or equivalently, nuclear corrections to
PDFs in nucleons, has been obtained \cite{EKR,EKS,EHKRS}.  There are,
however, some uncertainties in the 
determination of the initial conditions
for the $Q^2$ evolution due to the lack of experimental data. For
instance, gluons and sea quarks are not 
constrained \cite{EKST} in the EMC region. Also, it is a common belief that 
nuclear gluon distributions are
very weakly constrained by experimental data at small $x$.  The situation
would be similar to that of nucleon PDFs before HERA measured the
rapid increase of $F_2$ as $x\to 0$.  However, we 
have shown
\cite{EHKS} that within the 
leading twist (LT), lowest order (LO) DGLAP framework, the DIS data
provide enough constrains to rule out very strongly shadowed gluon
distributions in the region $x\gsim 0.01$.  The key experimental
information comes from the $Q^2$ dependence of the ratios
$F_2^{\rm Sn}/F_2^{\rm C}$ measured by the NMC Collaboration
\cite{NMC}.

In the case that the dominant contribution to total multiplicities
comes from jets or minijets, as has been proposed, the measurements of
$N_{ch}$ done at RHIC give direct information about the initial gluon
distributions in nuclei.
This has been used in Ref. \cite{newHIJING} to parametrize the gluon
distribution function in the HIJING model. As a result, a strong gluon
shadowing has been proposed in order to compensate the large
multiplicity obtained in the original model.

In previous works \cite{EKR,EKS} we have
determined a set of 
nPDFs (named EKS98) by 
using available data of $F_2^A(x,Q^2)$ and Drell-Yan  
cross sections in nuclear scatterings as constraints.
The iterative procedure is very similar to that in global fits of
parton distributions of the free proton. As a result the nPDFs become fixed
at an initial scale $Q_0^2$, and the evolution to $Q^2>Q_0^2$ is given by
the DGLAP equations \cite{dglap}.
The nuclear case is more complicated as
a new variable ($A$) is present. Moreover, as the amount of data does not
allow to fix the nuclear ratios for each parton species

\begin{equation}
R_{i}^A(x,Q^2)={f_i^A(x,Q^2)\over f_i(x,Q^2)},
\ \ \ \ \ i=g,\ u_v,\ d_v,\ \bar u,\ \dots ,
\label{eq1}
\end{equation}

\noindent
in an unambiguous way, 
some assumptions were imposed
in addition to the constraints from the momentum sum rule (that,
roughly speaking fixed the amount of antishadowing for gluons) and
baryon number conservation. 
The assumptions include the saturation of the shadowing at
small values of $x$ and the existence of an EMC effect for gluons and
sea quarks at large $x$.

Notice that the DGLAP evolution imposes some constraints
to nuclear gluons, just in the same way as it does
in the global analysis with nucleons. In
particular for small values of $x$, LO DGLAP equations  give

\begin{equation}
\frac{\partial R_{F_2}^A(x,Q^2)}{\partial \log Q^2}
\approx
\frac{10\alpha_s}{27\pi}\frac{xg(2x,Q^2)}{F_2^D(x,Q^2)}
\biggl\{R_G^A(2x,Q^2)-R_{F_2}^A(x,Q^2)\biggr\}.
\label{RF2slope}
\end{equation}

\noindent
So, the slope of the $Q^2$ evolution of the 
{\it ratio} $F_2^A/F_2^D$ is related to the difference of the nuclear ratio in
gluons and $F_2$ (notice, however, the factor $2x$ in $R_g$). As
$R_{F_2}^A(x,Q^2)$ is measured experimentally, data on the $Q^2$
evolution of the ratios $R_{F_2}^A$ constrain $R_g$ once
DGLAP evolution is taken into account. In particular, the experimental
data from the NMC Collaboration \cite{NMC} show that $F_2^{\rm
Sn}/F_2^{\rm C}$ increases as $Q^2$ increases at small values of
$x$. For this reason, $R_g^A$ cannot be much smaller that $R_{F_2}^A$
in order for the slope to be positive.  We
have studied different parametrizations for the initial
ratios,
and shown this fact by performing the DGLAP evolution without
making use of the approximation (\ref{RF2slope}). We will see that the
result is in agreement with the argument above.

The initial conditions of the four parametrizations studied can be
seen in Fig. 1. Notice that the HPC \cite{HPC} and the new--HIJING
\cite{newHIJING} parametrizations are $Q^2$ independent. They are 
used here as initial conditions for evolution.  HKM \cite{HKM} 
is based on  DGLAP evolution, 
but the NMC data \cite{NMC} on $Q^2$ dependence of $F_2^{\rm
Sn}/F_2^{\rm C}$ and DY data have not been taken into account.
We think that this is the main reason for the difference in
gluons and sea quarks as compared with EKS98.

\begin{figure}[tb]
\vspace{-0.5cm}
\centering{\epsfxsize=10cm\epsfbox{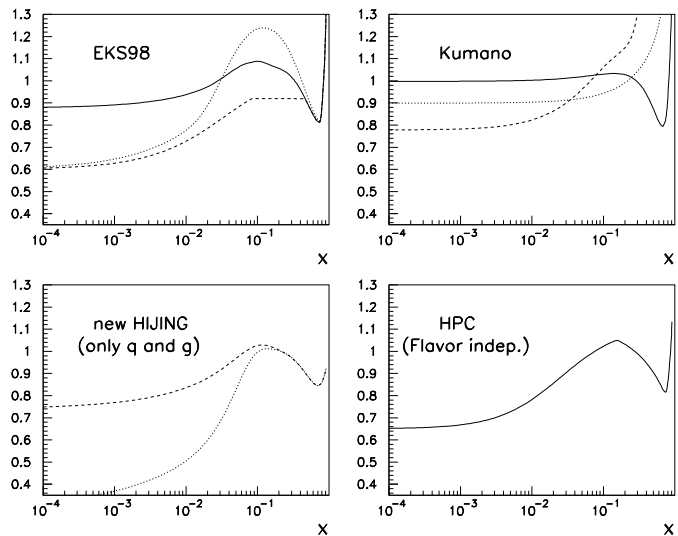}}
\vspace{0cm}
\caption[a]{
Initial ratios $R_i^A(x,Q_0^2)$ at $Q_0^2=2.25$ GeV$^2$ of Pb, for
valence quarks (solid), sea quarks (dashed) and gluons (dotted) from
the four different parametrizations used.}
\vspace{0cm}
\label{FigRgRF2}
\end{figure}

\begin{figure}[tb]
\vspace{-0.5cm}
\centering{\epsfxsize=12cm\epsfbox{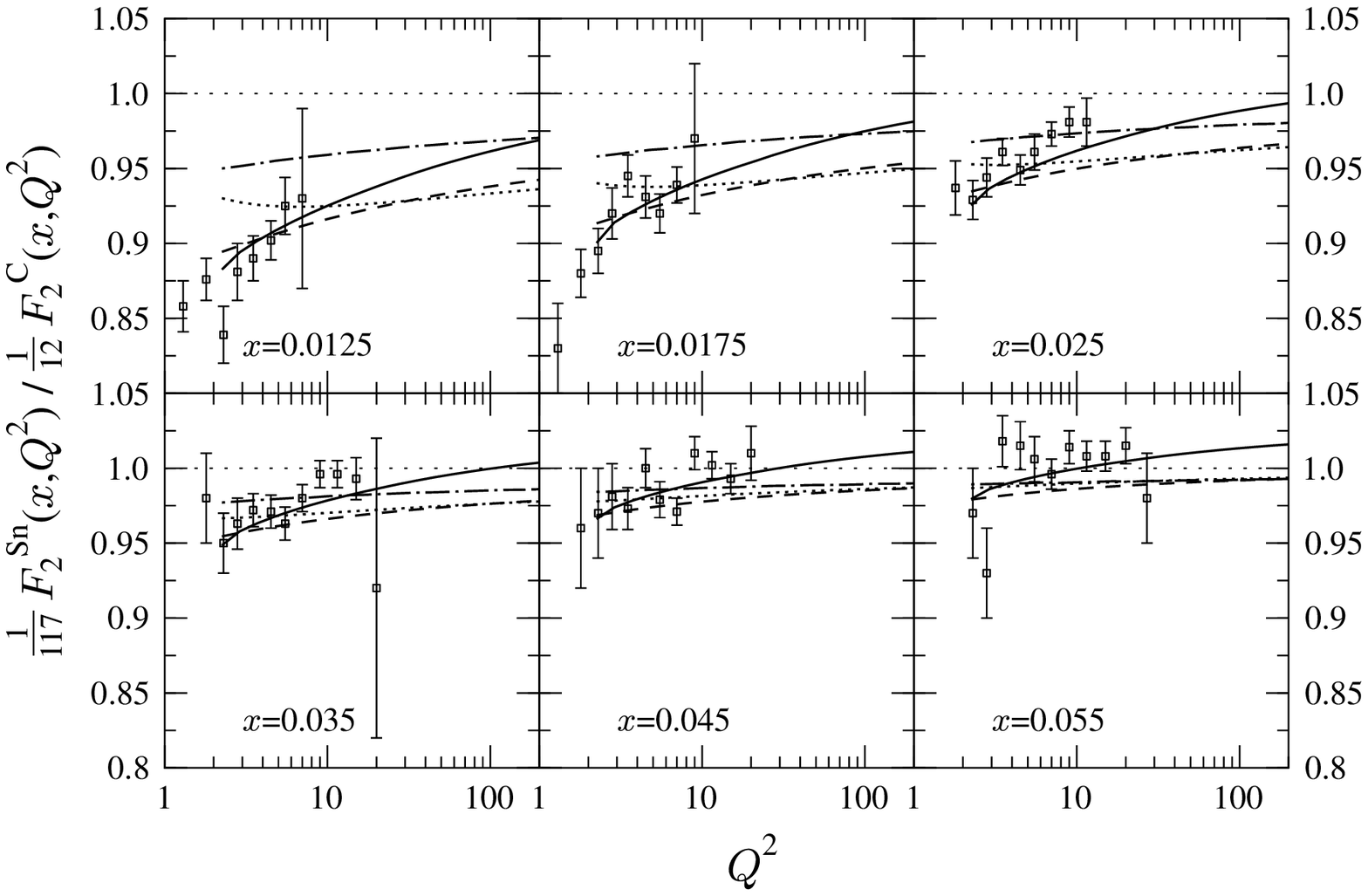}}
\vspace{-1cm}
\caption[a]{{ \small Comparison of the calculated \cite{EHKS} and 
measured $Q^2$
dependence of the ratio $F_2^{\rm Sn}/F_2^{\rm C}$. The NMC data
\cite{NMC} are shown with statistical errors only. The results for
EKS98 \cite{EKR,EKS} (solid lines) and HKM \cite{HKM}
(dotted-dashed) are from the corresponding global DGLAP analyses. The
$Q^2$ dependence of the  HPC \cite{HPC} (dashed) and HIJING \cite{newHIJING} 
(dotted) cases is obtained
from the DGLAP evolution with initial conditions similar to Fig. 1.
}}
\vspace{0cm}
\label{FigRF2}
\end{figure}

As can be seen from the figure, all the parametrizations give similar results
for quarks at the values of $x$ where experimental data exist.
However, the largest uncertainties are
in the gluon distribution function, the extreme case being the new
parametrization introduced in HIJING which 
suggests a very strong shadowing.
We will make use of this parametrization to demonstrate that a very
strong shadowing for gluons is ruled out 
based on the experimental data on the experimental data on the $Q^2$
dependence of $F_2^{\rm Sn}/F_2^{\rm C}$.  For that we perform 
the DGLAP evolution of the nPDFs given by these parametrizations for Sn
and C and, then, compute the ratios. In Fig. 2 we present the result
of this analysis. It is clear that the strong gluon shadowing case is
in complete disagreement with the data as it presents a negative
slope, while all the others are positive.

One could argue, however, that the non--linear terms (expected to be
present in evolution equations at high enough density) could change
this conclusion.  In order to check this possibility, we repeat the
$Q^2$ evolution using the same initial conditions but now taking into
account the gluon--fusion terms computed by Mueller and Qiu
\cite{MQ}. In this case, the evolution is
modified as can be seen in Fig. 3. Including these non--linear terms
the slopes become always smaller (even more negative in the case of
strong gluon shadowing). So we can conclude that using the presently
available experimental data, a strong gluon shadowing is ruled out at
$x\gsim 0.01$ in the present framework.

\begin{figure}[tb]
\vspace{-0.5cm}
\centerline{
\epsfxsize=12cm\epsfbox{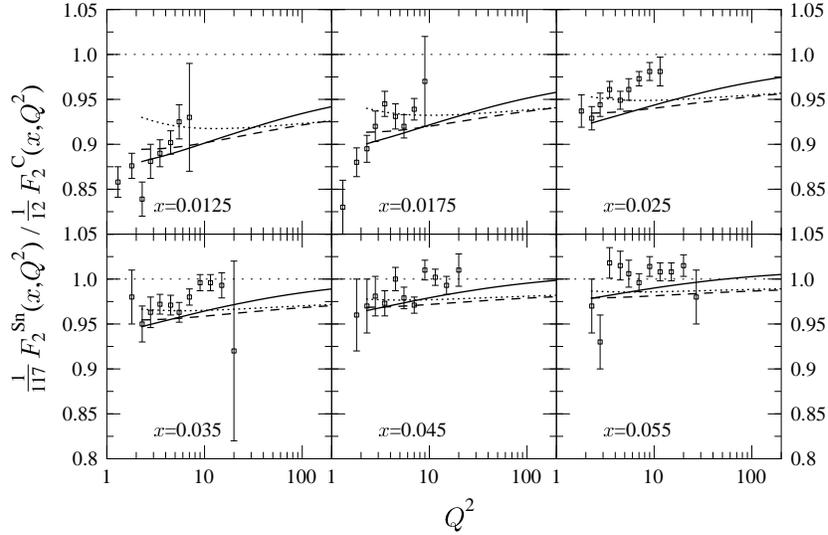}}
\vspace{-1cm}
\caption[a]{{ \small The scale dependence of the ratio $F_2^{\rm
   Sn}/F_2^{\rm C}$ calculated using DGLAP evolution with MQ
   corrections \cite{MQ}, and compared with the NMC data \cite{NMC}. Initial
   conditions for nuclear effects are taken from EKS98 (solid lines),
   HPC (dashed) and HIJING (dotted) parametrizations.   }}
\vspace{-0cm}
\label{FigRF2MQ}
\end{figure}

\section*{Acknowledgements}
C.A.S. is supported by a Marie Curie Fellowship of the European
Community programme TMR (Training and Mobility of Researchers), under
the contract number HPMF-CT-2000-01025.  Financial support from the
Academy of Finland, grant no. 50338, is gratefully acknowledged.


\begin{thebibliography}{99}
\bibitem{EKR}
  K.~J.~Eskola, V.~J.~Kolhinen and P.~V.~Ruuskanen,
  Nucl.\ Phys.\ B {\bf 535} (1998) 351.

\bibitem{EKS}
  K.~J.~Eskola, V.~J.~Kolhinen and C.~A.~Salgado,
  Eur.\ Phys.\ J.\ C {\bf 9} (1999) 61.

\bibitem{EHKRS}
  K.~J.~Eskola, H.~Honkanen, V.~J.~Kolhinen, P.~V.~Ruuskanen and C.~A.~Salgado,
  hep-ph/0110348.

\bibitem{EKST}
  K.~J.~Eskola, V.~J.~Kolhinen, C.~A.~Salgado and R.~L.~Thews,
  Eur.\ Phys.\ J.\ C {\bf 21} (2001) 613
  [hep-ph/0009251].

\bibitem{EHKS}
 K.~J.~Eskola, H.~Honkanen, V.~J.~Kolhinen and C.~A.~Salgado,
 Phys.\ Lett.\ B {\bf 532} (2002) 222
 [arXiv:hep-ph/0201256].

\bibitem{NMC}
  M.~Arneodo {\it et al.}  [New Muon Collaboration],
  Nucl.\ Phys.\ B {\bf 481} (1996) 23.

\bibitem{newHIJING}
  S.~y.~Li and X.~N.~Wang,
  Phys.\ Lett.\ B {\bf 527} (2002) 85
  [arXiv:nucl-th/0110075].

\bibitem{dglap}
  Yu. Dokshitzer, Sov. Phys. JETP {\bf 46} (1977) 1649;
  V.N. Gribov and L. N. Lipatov,
  Sov. Nucl. Phys. {\bf 15} (1972) 438, 675;
  G. Altarelli, G. Parisi, Nucl. Phys. B {\bf 126} (1977) 298.

\bibitem{HPC}
  Jan Czyzewski, K.J. Eskola, and J. Qiu, at the
  III International Workshop on Hard Probes of Dense Matter,
  ECT$^*$, Trento, June 1995.

\bibitem{HKM}
  M.~Hirai, S.~Kumano and M.~Miyama,
  Phys.\ Rev.\ D {\bf 64} (2001) 034003
  [hep-ph/0103208].

\bibitem{MQ}
  A.~H.~Mueller and J.~Qiu,
  Nucl.\ Phys.\ B {\bf 268} (1986) 427.

\end{thebibliography}
\end{document}